\documentclass[reprint,amsmath,amssymb,textcomp,aps]{revtex4-1}
\usepackage{graphicx}
\usepackage{dcolumn}
\usepackage{bm}

\DeclareMathOperator{\sgn}{sgn}

\begin{document}

\preprint{APS/123-QED}

\title{Excitons in planar quantum wells based on transition metal dichalcogenides}

\author{Pavel V. Ratnikov}
\email{ratnikov@lpi.ru}
\affiliation{A.M. Prokhorov General Physics Institute, Russian Academy of Sciences, ul. Vavilova 38, 119991 Moscow, Russia}

\date{\today}

\begin{abstract}
The problem of a size quantization for charge carriers in a planar quantum well consisting of different monolayers of transition metal dichalcogenides is solved using the Dirac model and the four-band model. For excitons, bound states of electrons and holes at the size quantization levels in such a quantum well, the energy spectrum was found in two cases: the Bohr radius is much smaller than the width of the quantum well (dielectric permeability of a substrate is relatively small) and the Bohr radius is much larger than it (the case of a strong dielectric screening). It is shown that the energy spectra in these two cases are completely different. A method for the synthesis of the heterostructures under consideration is also proposed.
\end{abstract}


\maketitle

\section{\label{s1}Introduction}

The solid state physics community is fascinated by two-dimensional (2D) materials. Great interest is caused by their unusual properties and the prospects provided by them in many areas, from nanoelectronics and photovoltaics to biological applications (e.g., biosensors or drug delivery). One of the brightest representatives of this rich diverse cohort is graphene, the most well studied to-date 2D crystal.

Starting from 2010s, different 2D materials have been used as the ``design cubes'' of vertical (layered) heterostructures. Transition metal dichalcogenides (TMDs) and their doped versions are particularly noteworthy for this purpose. Individual layers of 2D materials may be stacked on each other to synthesize single and double quantum wells (QWs), superlattices, etc. The layers are bound together through van der Waals attraction. Therefore, such heterostructures are also referred to as \emph{van der Waals heterostructures} \cite{GeimGrig}.

Given the number of different ways for stacking of 2D materials, it is possible to manufacture van der Waals heterostructures with any required properties. The inclusion of thin TMD layers in these heterostructures allows one to observe many-particle effects in systems with the long lifetimes charge carriers. At low temperatures, they may exhibit a superfluidity of excitons and superconductivity due to coupling of spatially separated quasiparticles \cite{Loz1, Loz2, Loz3, Loz4, Loz5, Loz6, Loz7} and condensation into an electron-hole liquid \cite{KS, And1, And2, Sil, And3}. Indirect excitons in van der Waals TMD-based heterostructures are recently studied in the work \cite{Cal}.

TMDs have a general chemical formula $MX_2$ with a transition metal atom $M$ usually from groups IV--VII (e.g., Hf, Nb, Ta, Mo, W, or Re) and two chalcogen atoms $X$ (S, Se, or Te). Their crystal structure was first established by Linus Pauling in 1923 \cite{Pau}. The monomolecular layer (monolayer) of TMD is a three-layer sandwich with a layer of metal atoms $M$ inserted between two layers of chalcogen atoms~$X$. Atoms in each layer are packed in a triangular lattice. Depending on the relative position of these layers, several types of structural phases are distinguished, mainly trigonal prismatic (2H) or octahedral (1T) phases. The 2H phases correspond to an ABA stacking when chalcogen atoms from different layers are located above each other. The 1T phases have an ABC stacking order. The thermodynamically stable phase is either the 2H or 1T phase. There also are the orthorhombic (distorted octahedral) 1T$_\text{d}$ and the monoclinic 1T$^\prime$ phases, which are often metastable ones \cite{Man}. For example, WTe$_2$ is undergoing the structure phase transition 1T$_\text{d}$$\rightarrow$1T$^\prime$ at high pressure \cite{Zhou, Lu}. The structure and synthesis of TMDs are described in more detail in the review \cite{Cher}.

By the end of the 1960s, about 60 TMDs were investigated, more than two thirds of which had a layered structure \cite{Wil}. Most of them are semiconductors with an indirect bandgap of $\sim$1 eV. The qualitative change occurs when going over from the bulk sample to the monolayer. It turned out that many 2D TMDs, including such well-known representatives as MoS$_2$, MoSe$_2$, WS$_2$, and WSe$_2$, become direct-band semiconductors with a bandgap of about 2 eV \cite{Mak1, Zhao, Zhang}.

Monolayers of TMDs have the conduction and valence-band extrema at the corners of the 2D hexagonal Brillouin zone \cite{Li, Leb}. Similar to graphene, there are two inequivalent valleys for low energy carriers. Since their intervalley scattering is suppressed, belonging to one of the two valleys (the valley index) may be considered a ``good'' quantum number. The usage of the valley degree of freedom in TMDs yields a promising option for a new type of nanoelectronics with the valley-selective charge carriers transport, called valleytronics. This is made possible by the valley-selective excitation of charge carriers with a circularly polarized electromagnetic wave~\cite{Xiao, Cao, Zeng, Mak2}.

We propose here a planar one-dimensional (1D) quantum well structure based on TMDs (Sec.~\ref{s2}). This paper is mainly devoted to two issues: the size quantization of charge carriers in such QWs (Sec.~\ref{s3}) and the energy spectrum of excitons in them depending on the dielectric environment (Sec.~\ref{s4}). These very straightforward questions are nevertheless very important for the physics of planar heterostructures composed of new 2D materials. In Sec.~\ref{s5} we discuss the possibilities to manufacture the TMD-based QWs \footnote{See Supplemental Material at ~~~~~~~~~~~~~~~~~~~~~~~~~~~~~~~~~~~~~~~~~~~~~~ for the schematic description of the manufacturing process of such heterostructures using the planar heterostructure MoTe$_2$/WTe$_2$/MoTe$_2$ growth as an example.} and summarize our results.

\section{\label{s2}Monolayer planar quantum well based on TMDs}

\begin{figure}[h!]
\begin{center}
\includegraphics[width=0.48\textwidth]{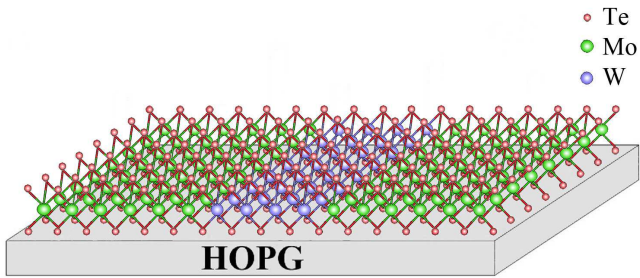}
\caption{\label{f1} (Color online) Schematic representation of the proposed here MoTe$_2$/WTe$_2$/MoTe$_2$ planar QW. Highly ordered pyrolytic graphite (HOPG) is used as a substrate.}
\end{center}
\end{figure}

We propose here a new type of  TMD-based planar heterostructures, namely MoSe$_2$/WTe$_2$/MoSe$_2$ or
MoTe$_2$/WTe$_2$/MoTe$_2$ single QWs. A schematic representation of the latter is given in Fig.~\ref{f1}.

Both QWs are examples of type~I QWs owing to the ratio of the bandgap $E_g$ and electron affinity $\chi$ for monolayer of MoSe$_2$ ($E_g=2.25$ eV \cite{Liu1} and $\chi=3.21$ eV~\footnote{This is a corrected value, because the authors of the paper \cite{Xen} have used a clearly underestimated value $E_g=1.58$ eV. For comparison: the optical bandgap of MoSe$_2$ is equal to $E^\text{opt}_g=1.659$ eV \cite{Ross}. The authors of the paper \cite{Xen} experimentally measured the difference between the edges of the valence bands of AlN and MoSe$_2$ (it is equal to 2.84 eV). In order to obtain the correct value of $\chi$, we took the value $E_g=2.25$ eV.}), WTe$_2$ ($E_g=1.18$ eV \cite{Kum1} and $\chi=3.69$ eV \cite{Gong}), and MoTe$_2$ ($E_g=1.72$ eV \cite{Yang} and $\chi=3.4$ eV~\cite{Ras}).

Such QWs can be synthesized as a result of varying of transition metal atoms in one plane. Although this is a rather complex approach, it brings a greater challenge from the technological side, which may push for further progress in the field of heterostructure synthesis.

In what follows we investigate theoretically the electron and hole size quantization and confined excitons in the proposed TMD-based planar QWs.

\section{\label{s3}Size quantization problem for charge carriers}

\subsection{\label{s3a}Dirac model}

The Dirac model is very constructive from the methodological side, allowing to obtain a rather simple dispersion relation for the size quantization levels. However, it is insufficient to describe the asymmetry of the dispersion of electrons and holes in the \textbf{K} valleys, since it automatically gives equal effective masses for them. This model does not take into account the absence of the center of inversion in the material. Considering these circumstances is necessary, for example, when analyzing the splittings of the spin levels of excitons in a magnetic field. The Dirac model leads to the same $g$-factors of the conduction band and the valence band, which, in turn, determines the absence of splittings of the spin levels of excitons in a magnetic field. However, the available experimental data show the presence of such splittings \cite{Dur}.

These features can be taken into account by including in the effective Hamiltonian the nearest in energy bands of the same parity, the bands $c+2$ and $v-3$ \cite{Kor}. Such a four-band Hamiltonian is presented in Subsec.~\ref{s3b}.

We emphasize that from the point of view of performing computations (numerical calculations), the Dirac model is also useful as the first iteration to find the size quantization levels. This makes it easier to find the right solutions within the four-band model.

\begin{figure}[b!]
\begin{center}
\includegraphics[width=0.48\textwidth]{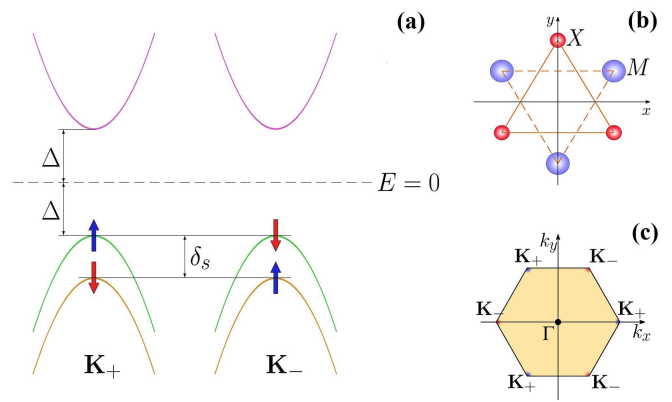}
\caption{\label{f2} (Color online) (a) The lower conduction band and the upper valence band at two valleys \textbf{K}$_+$ and \textbf{K}$_-$. The spin splitting of the conduction band is neglected, considering it as spin degenerate one, while the valence band has a strong spin splitting. (b) A top view of one section of a TMD crystal lattice with a coordinate reference (in the case of the heterostructure under consideration $M$\,=\,Mo,~W and $X$\,=\,Te). (c) The Brillouin zone of TMDs in the form of a regular hexagon with \textbf{K}$_+$ and \textbf{K}$_-$ points in the corners.}
\end{center}
\end{figure}

Often, in the Dirac model for TMDs, the lower valence band split by spin-orbit interaction is also taken into account. The effective Hamiltonian has the corresponding term, which is proportional to the spin operator $\widehat{s}_z$ \cite{Xiao}. Here, we write the Hamiltonian as \footnote{The last term is written in such a form as maintain the origin of the energy $E=0$ at the middle of the bandgap between the lower conduction band and the upper valence band, as shown in Fig.~\ref{f2}a.}
\begin{equation}\label{1}
\widehat{H}=\gamma_3{\boldsymbol\sigma}\widehat{\mathbf{p}}^\tau+\Delta\sigma_z+\left(\tau s_z-\frac{1}{2}\right)\delta_s\frac{1-\sigma_z}{2},
\end{equation}
where $\gamma_3$ is the band parameter, similar to Fermi velocity $v_F$ in graphene, $\widehat{\mathbf{p}}^\tau=(\tau\widehat{p}_x,\,\widehat{p}_y)$, $\widehat{p}_x=-i\partial_x$, and $\widehat{p}_y=-i\partial_y$ are components of the momentum operator ($\hbar=1$), $\tau=\pm1$ is the valley index ($\tau=+1$ for the valley \textbf{K}$_+$ and $\tau=-1$ for the valley \textbf{K}$_-$, see Fig.~\ref{f2}a, c), $\Delta=E_g/2$ is the half-width of the bandgap between the lower conduction band ($c$) and the upper valence band ($v$). The matrices $\sigma_x$, $\sigma_y$ and $\sigma_z$ are the Pauli matrices. The quantum number $s_z=\pm\frac{1}{2}$ is the eigenvalue of the spin operator $\widehat{s}_z$. The quantity $\delta_s$ is spin splitting at the valence band top caused by the spin-orbit interaction.

According to the results of first-principles calculations based on density functional theory, there are giant spin splittings from $\delta_s=148$ meV for MoS$_2$ to $\delta_s=456$ meV for WSe$_2$ \cite{Zhu} and $\delta_s=480$ meV for WTe$_2$ \cite{Zib}.

In our opinion, such a large splitting allows to omit the last term in (\ref{1}) in the framework of the two-band model and consider only the two nearest bands, namely, spin-polarized valence band with spin $\uparrow$ for $\tau=+1$ and spin $\downarrow$ for $\tau=-1$. Thus, we arrive at the 2$\times$2 effective Dirac Hamiltonian
\begin{equation}\label{2}
\widehat{H}^\tau_\text{D}=\gamma_3{\boldsymbol\sigma}\widehat{\mathbf{p}}^\tau+\Delta\sigma_z+V,
\end{equation}
where the scalar potential $V$ describes the possible displacement of the middle of the bandgap relative to the vacuum level $E_\text{vac}$ when we compare different TMDs.

The 2$\times$2 Dirac equation is
\begin{equation}\label{3}
\widehat{H}^\tau_\text{D}\Psi_\tau=E_\tau\Psi_\tau,\hspace{0.15cm}\Psi_\tau=
\begin{pmatrix}
  \psi^c_\tau \\
  \psi^v_\tau
\end{pmatrix},
\end{equation}
where the scalar envelope wave functions $\psi^c_\tau$ and $\psi^v_\tau$ describe states in the conduction band and the valence band, respectively. Such a description can be constructed by analogy with the description of states on two mutually penetrating triangular Bravais sublattices $A$ and $B$ of graphene. For the TMD crystal lattice of the 2H phase, we can also see two mutually penetrating triangular sublattices in layers of $X$ and $M$ atoms in a top view (see Fig.~\ref{f2}b). The valley index $\tau$ is written in the general case at energy as well. As will be shown below, asymmetry between valleys is present in an asymmetric QW, due to the explicit dependence of the energy of charge carriers on $\tau$. Note that there is no such dependence for symmetric QWs.

The 4$\times$4 Dirac Hamiltonian (\ref{2}) is similar to the Dirac Hamiltonian in quantum electrodynamics (QED) $\widehat{H}_\text{D}=c{\boldsymbol\alpha}\widehat{\mathbf{p}}+\beta\Delta+V$, where ${\boldsymbol\alpha}=\bigl(\begin{smallmatrix} O & {\boldsymbol\sigma} \\ {\boldsymbol\sigma} & O \end{smallmatrix}\bigr)$ and $\beta=\bigl(\begin{smallmatrix}I & O \\ O & -I \end{smallmatrix}\bigr)$ are the Dirac matrices ($O$ and $I$ are the zero and unit matrices, respectively). 4-vector of the current density in QED is $j_\mu=(\overline{\Psi}\gamma_0\Psi,\,c\overline{\Psi}{\boldsymbol\gamma}\Psi)$, where $\overline{\Psi}=\Psi^\dagger\gamma_0$ is the Dirac conjugate bispinor and $\gamma_0=\beta$ and ${\boldsymbol\gamma}=\gamma_0{\boldsymbol\alpha}=\bigl(\begin{smallmatrix} O & {\boldsymbol\sigma} \\ -{\boldsymbol\sigma} & O \end{smallmatrix}\bigr)$ are the Dirac $\gamma$-matrices in the standard representation. It is seen that $\widehat{H}_\text{D}$ transfers to $\widehat{H}^\tau_\text{D}$ after replacements $c\rightarrow\gamma_3$, ${\boldsymbol\alpha}\rightarrow{\boldsymbol\sigma}$, $\beta\rightarrow\sigma_z$, and $\widehat{\mathbf{p}}\rightarrow\widehat{\mathbf{p}}^\tau$ with a decrease in the dimensionality of the space from 3 to 2. Therefore, when we repeat the output of the expression for the current density operator as in QED, we get that the ``current density'' is expressed by $\mathbf{j}^\tau=\gamma_3\Psi^\dagger_\tau{\boldsymbol\sigma}\Psi_\tau$. The components of this vector $j^\tau_x=\gamma_3\left(\psi^{c*}_\tau\psi^v_\tau+\psi^{v*}_\tau\psi^c_\tau\right)$ and $j^\tau_y=-i\gamma_3\left(\psi^{c*}_\tau\psi^v_\tau-\psi^{v*}_\tau\psi^c_\tau\right)$ must be continuous when passing through the boundary between two materials, $\left.j^\tau_x\right|_L=\left.j^\tau_x\right|_R$ and $\left.j^\tau_y\right|_L=\left.j^\tau_y\right|_R$, i.e., $\left.\gamma_3\psi^{c*}_\tau\psi^v_\tau\right|_L=\left.\gamma_3\psi^{c*}_\tau\psi^v_\tau\right|_R$. Here, the indexes $L$ and $R$ denote belonging to the region to the left and to the right of the boundary, respectively. The last equality is ensured by performing equalities $\left.\sqrt{\gamma_3}\psi^c_\tau\right|_L=\left.\sqrt{\gamma_3}\psi^c_\tau\right|_R$ and $\left.\sqrt{\gamma_3}\psi^v_\tau\right|_L=\left.\sqrt{\gamma_3}\psi^v_\tau\right|_R$ or the equality
\begin{equation}\label{4}
\left.\sqrt{\gamma_3}\Psi_\tau\right|_L=\left.\sqrt{\gamma_3}\Psi_\tau\right|_R.
\end{equation}
The boundary condition (\ref{4}) is also established for $\Psi_\tau$ by integrating the Dirac equation (\ref{3}) in the vicinity of the interface between the media \cite{SS, RS}.

\begin{figure}[t!]
\begin{center}
\includegraphics[width=0.32\textwidth]{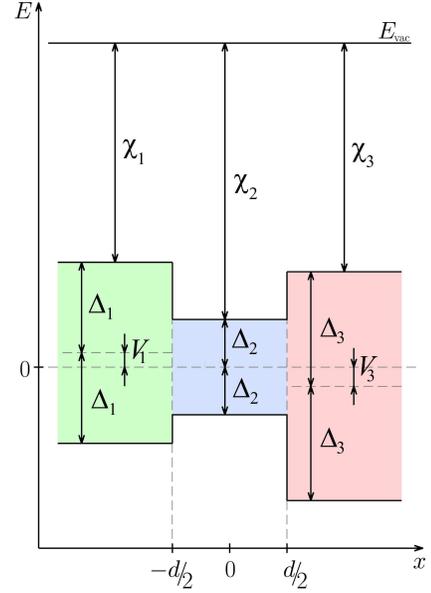}
\caption{\label{f3} (Color online) The energy diagram for QW under analysis: $E_\text{vac}$ is the vacuum level and $\chi_i$ ($i=1,\,2,\,3$) is the electron affinity.}
\end{center}
\end{figure}

Now, let us consider a QW. In the general case, we consider an asymmetric QW (e.g., MoTe$_2$/WTe$_2$/MoSe$_2$). Each region is characterized by numbers $\gamma_{3i}$, $\Delta_i$, and $V_i$ ($i=1,\,2,\,3$). Its energy diagram is shown schematically in Fig.~\ref{f3}. The $E=0$ level is set to coincide with the middle of the bandgap in the QW region, a strip of the TMD with a smaller bandgap, so that $V_2=0$. Then, the values of the scalar potential for the barrier regions are
\begin{equation}\label{5}
\begin{split}
   V_1&=\Delta_2+\chi_2-\left(\Delta_1+\chi_1\right),\\
   V_3&=\Delta_2+\chi_2-\left(\Delta_3+\chi_3\right),
\end{split}
\end{equation}
where $\chi_i$ is the electron affinity, i.e., a distance in energy of the edge of the conduction band to the vacuum level $E_\text{vac}$ (see also Fig.~\ref{f2}).

The $x$ axis is directed perpendicular to the QW interfaces (the orientation of the axes is shown in Fig.~\ref{f2}b). The width of the QW is $d$. We consider the boundaries between the materials as sharp. The solution to the Dirac equation (\ref{3}) in three regions is

1) $x<-d/2$
\begin{equation}\label{6}
\Psi_{\tau1}=C_1\begin{pmatrix}1 \\ \varkappa_{\tau1}\end{pmatrix}e^{k_1x+ik_yy},
\end{equation}
$\varkappa_{\tau1}=\frac{i\gamma_{31}(-\tau k_1+k_y)}{E_\tau+\Delta_1-V_1}$ and $E_\tau=V_1\pm\sqrt{\Delta^2_1+\gamma^2_{31}(k^2_y-k^2_1)}$;

2) $-d/2<x<d/2$
\begin{equation}\label{7}
\Psi_{\tau2}=C_2\begin{pmatrix}1 \\ \varkappa^+_{\tau2}\end{pmatrix}e^{i(k_2x+k_yy)}+\widetilde{C}_2\begin{pmatrix}1 \\ \varkappa^-_{\tau2}\end{pmatrix}e^{i(-k_2x+k_yy)},
\end{equation}
$\varkappa^\pm_{\tau2}=\frac{\gamma_{32}(\pm\tau k_2+ik_y)}{E_\tau+\Delta_2}$ and $E_\tau=\pm\sqrt{\Delta^2_2+\gamma^2_{32}(k^2_y+k^2_2)}$;

3) $x>d/2$
\begin{equation}\label{8}
\Psi_{\tau3}=C_3\begin{pmatrix}1 \\ \varkappa_{\tau3}\end{pmatrix}e^{-k_3x+ik_yy},
\end{equation}
$\varkappa_{\tau3}=\frac{i\gamma_{33}(\tau k_3+k_y)}{E_\tau+\Delta_3-V_3}$ and $E_\tau=V_3\pm\sqrt{\Delta^2_3+\gamma^2_{33}(k^2_y-k^2_3)}$.

Plus and minus in Eqs. (\ref{6})--(\ref{8}) the expression for the energy $E_\tau$ correspond to electrons and holes, respectively. The constants $C_1$, $C_2$, $\widetilde{C}_2$, and $C_3$ are found from the boundary condition (\ref{4}) and the normalization condition for wave functions (\ref{6})--(\ref{8})
\begin{equation}\label{9}
\int\limits_{-\infty}^\infty\Psi^\dagger_\tau\Psi_\tau dx=1.
\end{equation}

Using also the boundary condition (\ref{4}), we obtain that the carrier energy spectrum is determined by the following dispersion relation
\begin{equation}\label{10}
\tan\left(k_2d\right)=\frac{\tau A^-_{\tau}\gamma_{32}k_2}{A^+_{\tau}\gamma_{32}k_y-B_\tau\left(E_\tau+\Delta_2\right)-C_\tau\left(E_\tau-\Delta_2\right)},
\end{equation}
where
\begin{eqnarray*}
   A^\pm_\tau=&&\gamma_{31}\left(-\tau k_1+k_y\right)\left(E_\tau+\Delta_3+V_3\right) \\
   \pm&&\gamma_{33}\left(\tau k_3+k_y\right)\left(E_\tau+\Delta_1+V_1\right), \\
   B_\tau=&&\gamma_{31}\gamma_{33}\left(-\tau k_1+k_y\right)\left(\tau k_3+k_y\right), \\
   C_\tau=&&\left(E_\tau+\Delta_1+V_1\right)\left(E_\tau+\Delta_3+V_3\right).
\end{eqnarray*}

Due to the explicit dependence on $\tau$ in Eq.~(\ref{10}), the dispersion curve in one valley does not coincide with the dispersion curve in another valley, but they turn into each other when the sign of $k_y$ is changed. The valleys are connected via the time inversion transformation.

For the symmetric QW when $\gamma_{33}=\gamma_{31}$, $\Delta_3=\Delta_1$, and $V_3=V_1$ [the potential barrier on the right is the same as on the left and the system is symmetric with respect to the $x\rightarrow-x$ transformation], the explicit dependence on $\tau$ disappears and Eq.~(\ref{10}) is rewritten as
\begin{equation*}\label{10p}
\tan\left(k_2d\right)=\frac{\gamma_{31}\gamma_{32}k_1k_2}{E\left(E-V_1\right)-\Delta_1\Delta_2-\gamma^2_{31}k^2_y}.\tag{10$'$}
\end{equation*}

\subsection{\label{s3b}Four-band model}

As it was stated in the beginning of the subsection \ref{s3a}, the transition to the four-band model is carried out by adding bands of the same parity as the lower conduction band $c$ and the upper valence band $v$, and lying in energy in proximity to them: above $c$ there is $c+2$, below $v$ there is $v-3$ \cite{Dur}.

Let us work in the basis of wave functions $\left\{\left|\psi^{c+2}_\tau\right\rangle,\:\left|\psi^c_\tau\right\rangle,\:\left|\psi^v_\tau\right\rangle,\:\left|\psi^{v-3}_\tau\right\rangle\right\}$. The effective Hamiltonian 4$\times$4 has the form \cite{Wang}
\begin{equation}\label{11}
\widehat{H}^\tau_\text{4b}=
\begin{pmatrix}
  E_{c+2} & \gamma_6\widehat{p}^\tau_- & \gamma_4\widehat{p}^\tau_+ & 0 \\
  \gamma_6\widehat{p}^\tau_+ & E_c & \gamma_3\widehat{p}^\tau_- & \gamma_5\widehat{p}^\tau_+ \\
  \gamma_4\widehat{p}^\tau_- & \gamma_3\widehat{p}^\tau_+ & E_v & \gamma_2\widehat{p}^\tau_- \\
  0 & \gamma_5\widehat{p}^\tau_- & \gamma_2\widehat{p}^\tau_+ & E_{v-3}
\end{pmatrix}.
\end{equation}
Here, $\widehat{p}^\tau_\pm=\tau\widehat{p}_x\pm i\widehat{p}_y$ and $\gamma_2$, $\gamma_3$, $\gamma_4$, $\gamma_5$, and $\gamma_6$ are the band parameters. The band edges $E_{v-3}$, $E_v$, $E_c$, and $E_{c+2}$ are counted from the middle of the bandgap between $c$ and $v$ bands. For reasons of conformity with the Dirac model, we take $E_v=-\Delta_i+V_i$ and $E_c=\Delta_i+V_i$. Moreover, it is possible to put $V_2=0$ for QW region.

We also consider the boundaries between materials to be sharp, so that smooth potentials do not arise in the boundary regions, and the band parameters $\gamma_j$ ($j=2-6$) are constants in each medium up to the boundary. Therefore, the ``symmetrization'' of the Hamiltonian (\ref{11}) by the introduction of anticommutators $\gamma_j\widehat{p}^\tau_\pm\rightarrow\frac{1}{2}\left\{\gamma_j,\:\widehat{p}^\tau_\pm\right\}$ is not required so that it remains Hermitian \cite{SS}.

The equation for the four-component envelope wave function with Hamiltonian (\ref{11})
\begin{equation}\label{12}
\widehat{H}^\tau_\text{4b}\Psi_\tau=E_\tau\Psi_\tau
\end{equation}
gives for free charge carriers the dispersion relation $\det\left(H^\tau_\text{4b}-E_\tau\right)=0$ [$H^\tau_\text{4b}$ with $\widehat{p}^\tau_\pm\rightarrow k^\tau_\pm=\tau k_x\pm ik_y$] which is the equation on $E_\tau$ of the fourth power in quasimomentum $\mathbf{k}$
\begin{widetext}
\begin{equation}\label{13}
\begin{split}
   &\left(E_{c+2}-E_\tau\right)\left(E_c-E_\tau\right)\left(E_v-E_\tau\right)\left(E_{v-3}-E_\tau\right)-\left(E_v-E_\tau\right)\left(E_{v-3}-E_\tau\right)\gamma^2_6k^\tau_+k^\tau_-\\
   -&\left(E_{c+2}-E_\tau\right)\left(E_{v-3}-E_\tau\right)\gamma^2_3k^\tau_+k^\tau_--\left(E_c-E_\tau\right)\left(E_{v-3}-E_\tau\right)\gamma^2_4k^\tau_+k^\tau_--\left(E_{c+2}-E_\tau\right)\left(E_c-E_\tau\right)\gamma^2_2k^\tau_+k^\tau_-\\
   -&\left(E_{c+2}-E_\tau\right)\left(E_v-E_\tau\right)\gamma^2_5k^\tau_+k^\tau_-+\left(E_{v-3}-E_\tau\right)\gamma_3\gamma_4\gamma_6\left(k^\tau_+\right)^3+\left(E_{c+2}-E_\tau\right)\gamma_2\gamma_3\gamma_5\left(k^\tau_+\right)^3\\
   +&\left(E_{v-3}-E_\tau\right)\gamma_3\gamma_4\gamma_6\left(k^\tau_-\right)^3+\left(E_{c+2}-E_\tau\right)\gamma_2\gamma_3\gamma_5\left(k^\tau_-\right)^3+\left(\gamma_2\gamma_6-\gamma_4\gamma_5\right)^2\left(k^\tau_+k^\tau_-\right)^2=0.
\end{split}
\end{equation}
\end{widetext}

In the quadratic in momentum approximation for electrons $E_\tau\approx E_c+\frac{k^\tau_+k^\tau_-}{2m^*_c}$ and for holes $E_\tau\approx E_v-\frac{k^\tau_+k^\tau_-}{2m^*_v}$, we obtain from equation (\ref{13}) the expressions for the effective mass of electrons $m^*_c$ and holes $m^*_v$ \cite{Wang}
\begin{equation}\label{14}
\begin{split}
\frac{1}{m^*_c}&=2\left[\frac{\gamma^2_5}{E_c-E_{v-3}}+\frac{\gamma^2_3}{E_c-E_v}+\frac{\gamma^2_6}{E_c-E_{c+2}}\right],\\
\frac{1}{m^*_v}&=2\left[\frac{\gamma^2_5}{E_{v-3}-E_v}+\frac{\gamma^2_3}{E_c-E_v}+\frac{\gamma^2_6}{E_{c+2}-E_v}\right].
\end{split}
\end{equation}
It can be seen that $m^*_v\neq m^*_c$.

Eliminating the wave function components $\psi^{c+2}_\tau$ and $\psi^{v-3}_\tau$ in equation (\ref{12}), we arrive at an effective Hamiltonian that takes into account the influence of the $c+2$ and $v-3$ bands
\begin{equation}\label{15}
\widehat{H}^\tau=\widehat{H}^\tau_\text{D}+\delta\widehat{H}^\tau,
\end{equation}
where
\begin{eqnarray*}
\delta\widehat{H}^\tau&=&
\begin{pmatrix}
  A_{56}\widehat{p}^\tau_+\widehat{p}^\tau_- & B^{25}_{46}\widehat{p}^\tau_+\widehat{p}^\tau_+ \\
  B^{25}_{46}\widehat{p}^\tau_-\widehat{p}^\tau_- & A_{24}\widehat{p}^\tau_+\widehat{p}^\tau_-
\end{pmatrix},\\
A_{ij}&=&\frac{\gamma^2_i}{E_\tau-E_{v-3}}-\frac{\gamma^2_j}{E_{c+2}-E_\tau}\hspace{0.1cm}(i=2,\,5;\:j=4,\,6),\\
B^{25}_{46}&=&\frac{\gamma_2\gamma_5}{E_\tau-E_{v-3}}-\frac{\gamma_4\gamma_6}{E_{c+2}-E_\tau}.
\end{eqnarray*}

In the quadratic in momentum approximation, the equation $\widehat{H}^\tau\Psi_\tau=E_\tau\Psi_\tau$ with the Hamiltonian (\ref{15}) for the wave function $\Psi_\tau=\bigl(\begin{smallmatrix}\psi^c_\tau \\ \psi^v_\tau\end{smallmatrix}\bigr)$ can be reduced to two equations separately for the functions $\psi^{c,v}_\tau$
\begin{equation}\label{16}
\left(\frac{1}{2m^*_{c,v}}\widehat{p}^\tau_+\widehat{p}^\tau_-+E_{c,v}\right)\psi^{c,v}_\tau=E_\tau\psi^{c,v}_\tau.
\end{equation}
The effective masses are given by Eqs. (\ref{14}).

The equations (\ref{16}) are second-order differential equations, so additional boundary conditions are needed that are different from (\ref{4}). They must ensure, as in the case of the usual Hamiltonian in the Schr\"{o}dinger equation, the continuity of the current density through the boundary between two materials for electrons $j^e_x=\frac{-i}{2m^*_c}\left(\psi^{c*}\partial_x\psi^c-\psi^c\partial_x\psi^{c*}\right)$ and for holes
$j^h_x=\frac{-i}{2m^*_v}\left(\psi^{v*}\partial_x\psi^v-\psi^v\partial_x\psi^{v*}\right)$. This is achieved with the continuity of $\psi^c_\tau$ and $\psi^v_\tau$ and combinations $m^{*-1}_c\partial_x\psi^c_\tau$ and $m^{*-1}_v\partial_x\psi^v_\tau$, which is analogous to the boundary condition used in \cite{Ben} and generalized by Bastard \cite{Bas1, Bas2}.

Thus, we can solve the QW size quantization problem for electrons with the wave function $\psi^c_\tau$ and for holes with the wave function $\psi^v_\tau$, satisfying Eqs. (\ref{17}), using the following boundary conditions
\begin{equation}\label{17}
\left.\psi^{c,v}_\tau\right|_L=\left.\psi^{c,v}_\tau\right|_R,\hspace{0.1cm}\left.\frac{1}{m^*_{c,v}}\partial_x\psi^{c,v}_\tau\right|_L=\left.\frac{1}{m^*_{c,v}}\partial_x\psi^{c,v}_\tau\right|_R.
\end{equation}

It should be noted that the valley index $\tau$ disappears from the equation (\ref{17}): $\widehat{p}^\tau_+\widehat{p}^\tau_-\equiv\widehat{p}^2_x+\widehat{p}^2_y$ ($\tau^2=1$). Thus, the four-band model reduced to Eq. (\ref{16}) does not take into account possible valley asymmetry of dispersion curves corresponding to size quantization levels, but the electron-hole asymmetry is clearly taken into account. This is more important for finding the exciton energy spectrum. In what follows, we omit the $\tau$ index of wave functions and energy.

Now, let us get the dispersion relation for the size quantization levels in the QW. For definiteness, let us consider the case of electrons and characterize each region of QW by numbers $E_{ci}$ and $m^*_{ci}$ ($i=1,\,2,\,3$) [for holes, the energy sign changes and $c\rightarrow v$]. The solution of Eq. (\ref{16}) in three regions is

1) $x<-d/2$
\begin{equation}\label{18}
\psi^c=c_1e^{k_1x+ik_yy},
\end{equation}
\begin{equation}\label{18p}
E=E_{c1}+\frac{1}{2m^*_{c1}}\left(k^2_y-k^2_1\right);\tag{18$'$}
\end{equation}

2) $-d/2<x<d/2$
\begin{equation}\label{19}
\psi^c=c_2e^{i(k_2x+k_yy)}+\widetilde{c}_2e^{i(-k_2x+k_yy)},
\end{equation}
\begin{equation}\label{19p}
E=E_{c2}+\frac{1}{2m^*_{c2}}\left(k^2_y+k^2_2\right);\tag{20$'$}
\end{equation}

3) $x>d/2$
\begin{equation}\label{20}
\psi^c=c_3e^{-k_3x+ik_yy},
\end{equation}
\begin{equation}\label{20p}
E=E_{c1}+\frac{1}{2m^*_{c1}}\left(k^2_y-k^2_3\right).\tag{20$'$}
\end{equation}

The constants $c_1$, $c_2$, $\widetilde{c}_2$, and $c_3$ are found from the normalization condition for wave functions (\ref{18})--(\ref{20}) similar to Eq. (\ref{9}). Matching the wave functions at the QW boundaries $x=-d/2$ and $x=d/2$, we obtain the dispersion relation for electrons on the size quantization levels
\begin{equation}\label{21}
\tan\left(k_2d\right)=k_2\frac{m^*_{c1}k_3+m^*_{c3}k_1}{\widetilde{m}^*_ck^2_2-m^*_{c2}k_1k_3},\hspace{0.1cm}\widetilde{m}^*_c\equiv\frac{m^*_{c1}m^*_{c3}}{m^*_{c2}}.
\end{equation}

Eliminating $k_1$ and $k_3$ from Eq. (\ref{21}) using Eqs. (\ref{18p})--(\ref{20p}), we can find the function $k_2(k_y)$ and, consequently, the energy $E_{N_e}(k_y)$ for each $N_e$th size quantization level according to Eq. (\ref{19p}). Since the valley asymmetry is absent, the extremum of all dispersion curves $E_{N_e}(k_y)$ lies at $k_y=0$, i.e., at \textbf{K}$_+$ or \textbf{K}$_-$ point in the Brillouin zone. The first derivative of the function $k_2(k_y)$ at the point $k_y=0$ is equal to zero, $k^\prime_{20}=k^\prime_2(k_y=0)=0$. The same is true for holes. The effective mass of electrons on the $N_e$th size quantization level is given by
\begin{equation}\label{22}
\frac{1}{m^*_c}=\left.\frac{\partial^2E_{N_e}}{\partial k^2_y}\right|_{k_y=0}=\frac{1+k_{20}k^{\prime\prime}_{20}}{m^*_{c2}},
\end{equation}
where $k_{20}=k_2(k_y=0)$ and $k^{\prime\prime}_{20}=k^{\prime\prime}_2(k_y=0)$ are values of the function $k_2(k_y)$ and its second derivative at the point $k_y=0$.

For a symmetric QW [$E_{c3}=E_{c1}$ and $m^*_{c3}=m^*_{c1}$], Eq.~(\ref{21}) is reduced to
\begin{equation}\label{21p}
\tan\left(k_2d\right)=\frac{k_1k_2}{\kappa k^2_2-m^*_{c2}U_0},\hspace{0.1cm}\kappa\equiv\frac{m^*_{c1}+m^*_{c2}}{2m^*_{c2}},\tag{21$'$}
\end{equation}
where $U_0=E_{c1}-E_{c2}$ is the height of potential barriers. Eq. (\ref{21}$^\prime$) is equivalent to the equation (3) in the solution of the problem 2 after \textsection\,22 of the book \cite{LL}, when the effective masses $m^*_{c1}$ and $m^*_{c2}$ are the same and $\kappa=1$.

\begin{figure}[t!]
\begin{center}
\includegraphics[width=0.3549\textwidth]{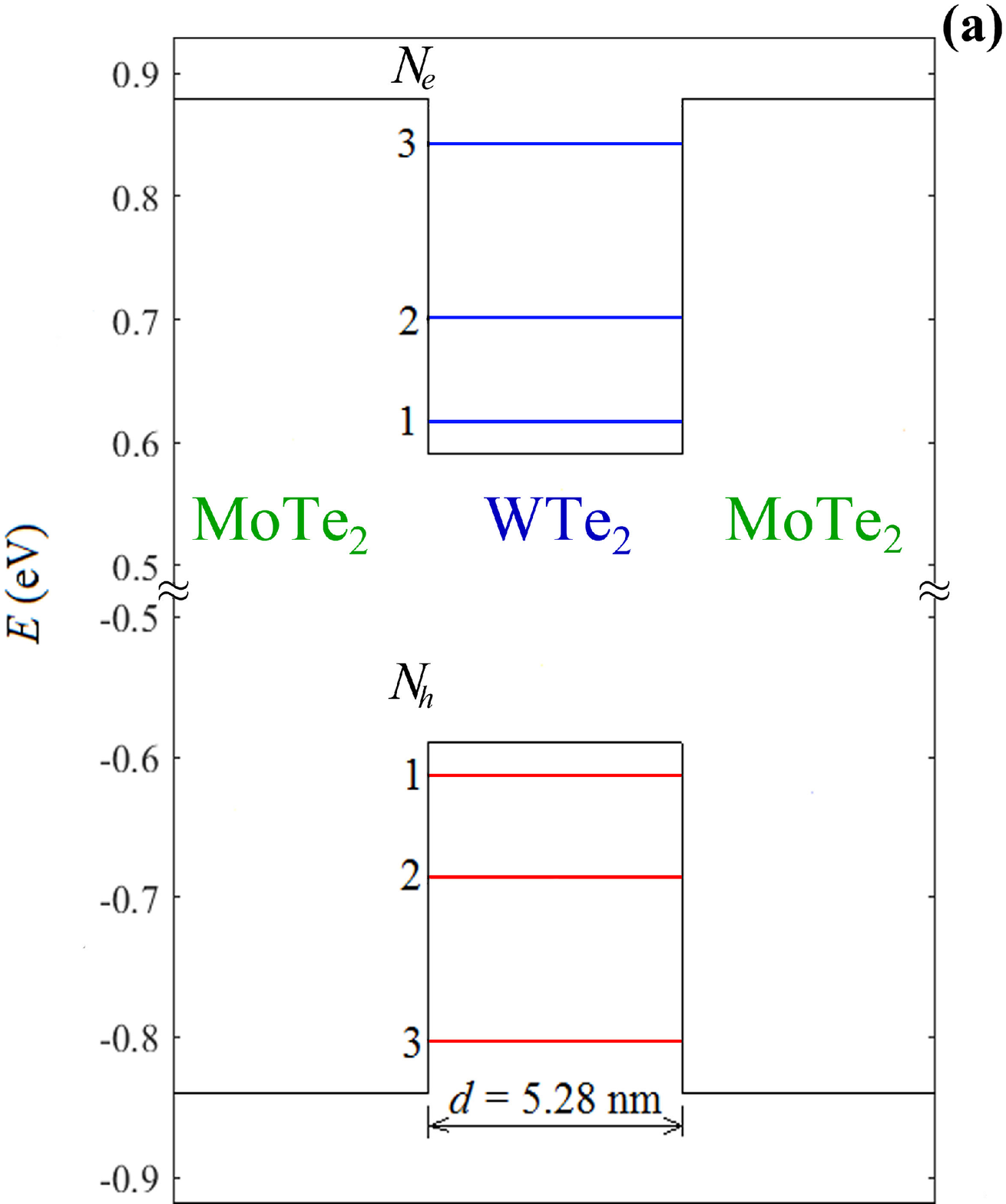}

~

\includegraphics[width=0.41\textwidth]{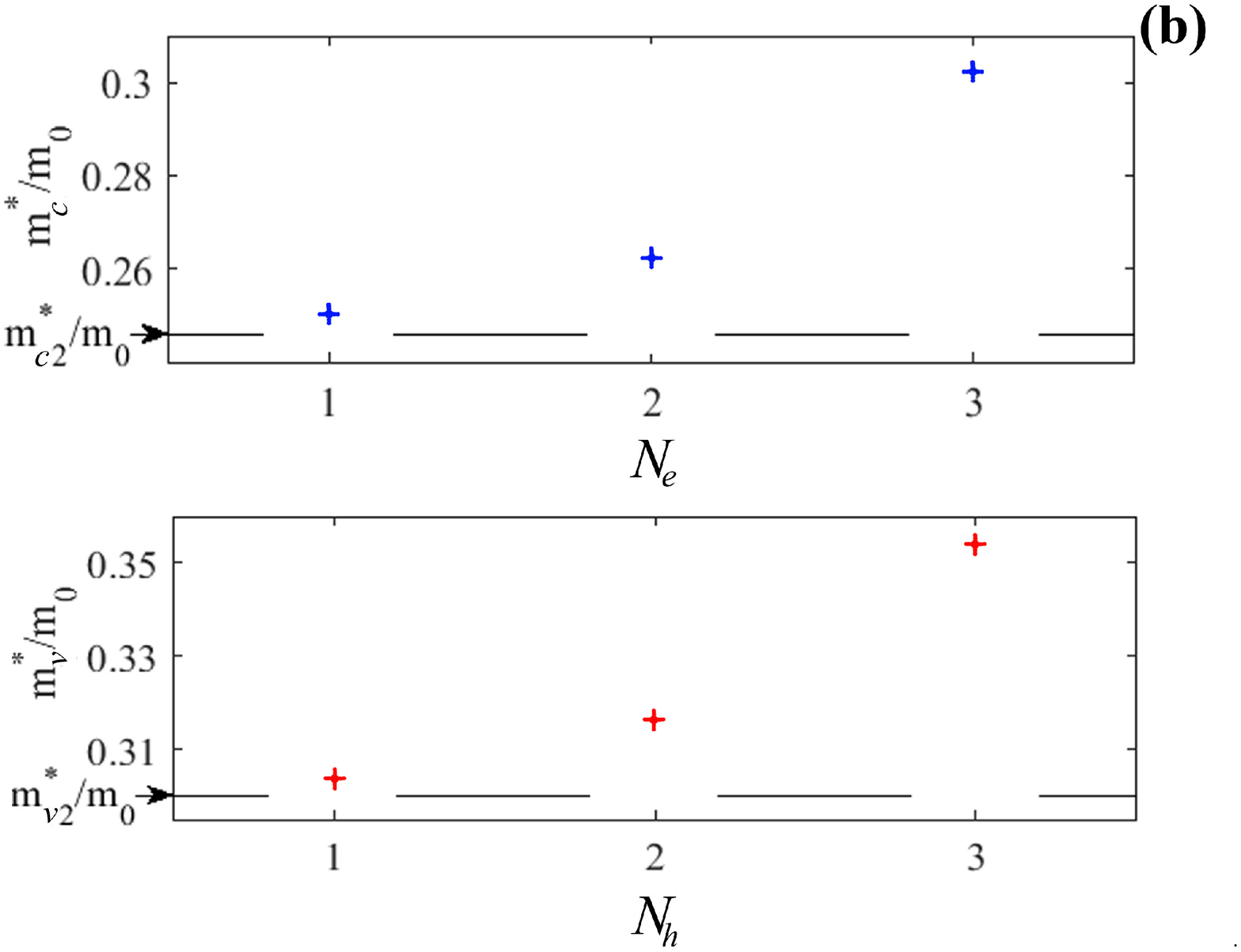}
\caption{\label{f4} (Color online) The results of numerical calculations for the MoTe$_2$/WTe$_2$/MoTe$_2$ heterostructure. (a) Values of the energy for size quantization levels of electrons ($E^e_1=616.5$ meV, $E^e_2=700.8$ meV, $E^e_3=842.8$ meV) and of holes ($E^h_1=-613.6$ meV, $E^h_2=-685.9$ meV, $E^h_3=-803.4$ meV). (b) Values of the effective mass of electrons ($m^*_c/m_0=0.25,\,0.262,\,0.302$) [the upper panel] and of holes ($m^*_v/m_0=0.304,\,0.316,\,0.354$) [the lower panel] at the extremes of the corresponding dispersion curves.}
\end{center}
\end{figure}

As an example, let us calculate the size quantization levels in MoTe$_2$/WTe$_2$/MoTe$_2$ QW with the values of parameters $E_{ci}$ and $E_{vi}$ obtained from the ratio of the bandgaps $E_{gi}$ and the electron affinity $\chi_i$ ($i=1$ for MoTe$_2$, $i=2$ for WTe$_2$) presented in Section~\ref{s2}. The height of potential barriers for electrons is $U^e_0=\chi_2-\chi_1=290$ meV, and for holes is $U^h_0=\chi_1+E_{g1}-\left(\chi_2+E_{g2}\right)=250$ meV. The effective masses of electrons and holes are $m^*_{c1}=0.655m_0$, $m^*_{c2}=0.246m_0$ and $m^*_{v1}=0.618m_0$, $m^*_{v2}=0.3m_0$ ($m_0$~is~the free electron mass) \cite{Zib}. The QW width $d$ is taken as a multiple of the lattice constant $b$ (the distance between neighboring tellurium atoms in one layer), $b=3.52$ \AA\, \cite{Far}. We take $d=15b=5.28$ nm. Using Eq.~(\ref{21p}) for electrons and its analogue for holes, we determine three electron levels and three hole levels inside the QW [$E^e_{N_e}-E_{c2}<U^e_0$ and $E_{v2}-E^h_{N_h}<U^h_0$] (see Fig.~\ref{f4}a). Using the formula (\ref{22}) for the electron effective mass and its analogue for holes, we find the corresponding effective masses. Note that with an increase in the size quantization level number, the effective mass increases for both electrons and holes (see Fig.~\ref{f4}b).

Similarly, one can find size quantization levels and the corresponding effective masses in the potential well for holes in the valence band split off by spin-orbit interaction. For this, one should substitute the effective hole masses $m^*_{v1}$ and $m^*_{v2}$ of the split off valence band into an equation similar to equation (\ref{21p}).

\section{\label{s4}Excitons}

A striking feature of the excitons in monolayers of TMDs is their large binding energy and small Bohr radius in the ground state (the $1s$ state). Typical values are $|E_{1s}|\simeq500$ meV and $a_1\simeq10$ \AA\, for freely suspended films in vacuum \cite{Dur}.

Two series of peaks are often observed in the photoluminescence spectrum of TMD monolayers due to a large spin splitting of the valence band, usually named as $A$ and $B$. The peak $A$ corresponds to the exciton which is binding state of an electron in the conduction band $c$ and a hole in the upper valence band $v$, while the peak $B$ corresponds to the exciton with a hole in the valence band split off by the magnitude of the spin splitting $\delta_s$ (see Fig.~\ref{1}a). The peak $B$ has a blue shift relative to the peak $A$.

The additional advantage of WTe$_2$ in the QW region is the largest valence band spin splitting among the TMD monolayers, $\delta_s=480$ meV \cite{Zib}. Thus, the energy distance between peaks $A$ and $B$ will also be the largest in MoTe$_2$/WTe$_2$/MoTe$_2$ QW. Moreover, $\delta_s>U^{e,h}_0$. This makes it possible to excite only $A$ peak when the frequency interval of the exciting laser $\omega_\text{min}<\omega<\omega_\text{max}$ is chosen so that $\omega_\text{max}-\omega_\text{min}<\delta_s$ and $E_g(\text{WTe}_2)<\omega_\text{max}<E_g(\text{MoTe}_2)$, e.g., $\omega_\text{min}=E^e_1-E^h_1$ and $\omega_\text{max}=E^e_3-E^h_3$ for the example considered at the end of Section~\ref{s2}. Below, we focus only on the $A$ exciton energy spectrum. Although it will become clear from the foregoing that the calculation of the $B$ exciton energy spectrum is completely analogous if an effective mass of holes is found in the band split off by the spin.

We consider the planar QW as the monolayer film system on the substrate. The Bohr radius of the exciton $a_1$ will always be greater than its value for a suspended film. However, unlike large samples of the TMD monolayers, we have an additional characteristic scale of distances in QW, its width $d$. Therefore, two cases should be distinguished: (i) a weak dielectric screening, when $a_1\ll d$ (e.g., in the case of the SiO$_2$ substrate); (ii)~a strong dielectric screening, when $a_1\gg d$ (e.g., in~the~case~of the TiO$_2$ substrate).

\subsection{\label{s4a}Weak dielectric screening}

The presented above typical values of the binding energy $|E_{1s}|$ and of the Bohr radius $a_1$ support the applicability of a description of the exciton in the TMD films by the smooth envelope functions method, when the exciton wave function covers a large number of crystal unit cells \cite{Dur}.

Since the ``size'' of the excitons is assumed to be much smaller than the width of the QW, the motion of the electron and hole will be quasi-2D in the WTe$_2$ stripe, neglecting the charge carrier motion along the $z$ axis.

The Hamiltonian describing the 2D relative electron-hole motion in the exciton is
\begin{eqnarray}
\widehat{H}_\text{ex}&=&\widehat{T}+\widehat{U},\label{23}\\
\widehat{T}&=&\frac{1}{2\mu^*}\left(\frac{\partial^2}{\partial\rho^2}+\frac{1}{\rho}\frac{\partial}{\partial\rho}-\frac{l^2}{\rho^2}\right),\label{24}\\
\widehat{U}&=&-\frac{\pi\widetilde{e}^2}{2r^\prime_0}\left[H_0\left(\frac{\rho}{r^\prime_0}\right)-Y_0\left(\frac{\rho}{r^\prime_0}\right)\right].\label{25}
\end{eqnarray}
Here, $\mu^*$ is the reduced mass of the electron and hole, $\mu^{*-1}=m^{*-1}_c+m^{*-1}_v$, and $\rho=\left|{\boldsymbol\rho}_e-{\boldsymbol\rho}_h\right|$ is the distance between the electron and hole in the plane $z=0$, ${\boldsymbol\rho}_{e,h}=(x_{e,h},\,y_{e,h},\,0)$. The quantum number $l$ is the angular momentum, $l=0,\,1,\,2,\,\ldots$ We introduced the notation $\widetilde{e}^2=e^2/\varepsilon_\text{eff}$, where $\varepsilon_\text{eff}=(\varepsilon_1+\varepsilon_2)/2$ is the effective dielectric constant ($\varepsilon_1$ and $\varepsilon_2$ are values of the dc permittivity of the
materials above and below the film, respectively) \cite{Loz8, Kel}. Quantity $r^\prime_0=r_0/\varepsilon_\text{eff}$ and $r_0=2\pi\alpha_\text{2D}$, and $\alpha_\text{2D}$ is the 2D susceptibility of the QW region material (in our case, this is WTe$_2$), which can be estimated as $\widetilde{\alpha}_\text{2D}=L_c(\varepsilon_\perp-1)/4\pi$ with the interlayer separation between two chalcogen atoms layers $L_c$ and the in-plane component of the dielectric tensor $\varepsilon_\perp$~\cite{Ber}. As a rule, in comparison with $\alpha_\text{2D}$, obtained in calculations using density functional theory, this estimate is an estimate from above, i.e., $\alpha_\text{2D}\lesssim\widetilde{\alpha}_\text{2D}$. The functions $H_0$ and $Y_0$ are the Struve function and the Bessel function of the second kind (the Neumann function), respectively. The potential (\ref{25}) was derived by Keldysh \cite{Kel}.

To calculate the energy spectrum of the exciton, we use the variational approach. The trial wave function is taken in the form of the eigenfunctions of a 2D hydrogen atom \cite{PP}
\begin{eqnarray}\label{26}
\widetilde{\psi}_{nl}(\rho)&=&\frac{C_{nl}}{a}\left(\frac{2\rho}{a}\right)^le^{-\rho/a}L^{2l}_{n-l-1}\left(\frac{2\rho}{a}\right),\\
C_{nl}&=&\sqrt{\frac{(n-l-1)!}{\pi(n-\hspace{-0.05cm}^1\hspace{-0.08cm}/\hspace{-0.065cm}_2)(n+l-1)!}},\nonumber
\end{eqnarray}
where $n=1,\,2,\,\ldots$ is the principal quantum number, $0\leq l\leq n-1$, $a$ is the variational parameter, and $L^\alpha_\beta$ are the associated Laguerre polynomials.

Wave functions (\ref{26}) form a complete orthonormal set
\begin{equation*}
\int d^2\rho\widetilde{\psi}^*_{nl}(\rho)\widetilde{\psi}_{n^\prime l^\prime}(\rho)=\delta_{nn^\prime}\delta_{ll^\prime}.
\end{equation*}

\begin{figure*}[t!]
\begin{center}
\includegraphics[width=0.685\textwidth]{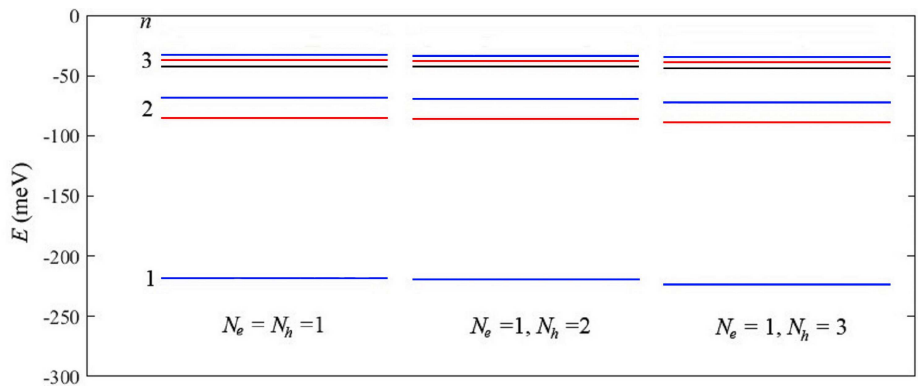}
\caption{\label{f5} (Color online) Three groups of levels for three excitons in the QW MoTe$_2$/WTe$_2$/MoTe$_2$ on the silicon dioxide substrate: for the exciton formed by the electron and the hole with $N_e=N_h=1$ (left), with $N_e=1$ and $N_h=2$ (center), and with $N_e=1$ and $N_h=3$ (right). The $s$ levels ($l=0$) are marked in blue, the $p$ levels ($l=1$) are red, and the $d$ levels ($l=2$) are black.}
\end{center}
\end{figure*}

When normalizing the wave functions (\ref{26}), we used the expression for the following integral \cite{Arf}
\begin{equation*}
\int\limits_0^\infty x^{2l+1}e^{-x}\left(L^{2l}_{n-l-1}(x)\right)^2dx=\frac{(n+l-1)!}{(n-l-1)!}(2n-1).
\end{equation*}

Wave functions (\ref{26}) also qualitatively reproduce the behavior of the exciton wave function obtained by more complex methods, for example, the solution of the Bethe-Salpeter equation (see Fig. 3b-e in \cite{Ye}). This confirms the applicability of $\widetilde{\psi}_{nl}(\rho)$ as trial wave functions.

The exciton energy is calculated as the average $\langle n,\,l|\widehat{H}_\text{ex}|n,\,l\rangle$ for the trial wave functions (\ref{26}) and depends on the variation parameter $a$
\begin{equation}\label{27}
E_{nl}(a)=\langle n,\,l|\widehat{T}|n,\,l\rangle+\langle n,\,l|\widehat{U}|n,\,l\rangle.
\end{equation}

It is easy to verify that the average kinetic energy operator (\ref{24}) for arbitrary $n$ and $l$ is equal to
\begin{equation}\label{28}
\langle n,\,l|\widehat{T}|n,\,l\rangle=\frac{1}{2\mu^*a^2}.
\end{equation}

The second term  on the right-hand side of Eq. (\ref{27}) is
\begin{equation}\label{29}
\begin{split}
&\langle n,\,l|\widehat{U}|n,\,l\rangle=-\frac{(n-l-1)!}{(2n-1)(n+l-1)!}\frac{\pi\widetilde{e}^2}{2r^\prime_0}\\
&\times\int\limits_0^\infty x^{2l+1}e^{-x}\left(L^{2l}_{n-l-1}(x)\right)^2\left[H_0(\varkappa x)-Y_0(\varkappa x)\right]dx.
\end{split}
\end{equation}
Hereinafter, $\varkappa=a/2r^\prime_0$.

The equation for the value of $a$, which corresponds to the minimum of the energy $E_{nl}(a)$, is
\begin{widetext}
\begin{equation}\label{30}
-\frac{1}{\mu^*a^3}-\frac{(n-l-1)!}{(2n-1)(n+l-1)!}\frac{\pi\widetilde{e}^2}{4r^{\prime2}_0}\int\limits_0^\infty x^{2(l+1)}e^{-x}\left(L^{2l}_{n-l-1}(x)\right)^2\left[\frac{2}{\pi}-H_1(\varkappa x)+Y_1(\varkappa x)\right]dx=0
\end{equation}
\end{widetext}
under the condition $\partial^2E_{nl}(a)/\partial a^2>0$.

We calculate the first $n$ from 1 to 3 levels of three $A$ excitons formed by the electron and the hole on the size quantization levels $N_e=1$ and $N_h=1$, $N_e=1$ and $N_h=2$, $N_e=1$ and $N_h=3$ (see Fig.~\ref{f5}). We used the estimate $\widetilde{\alpha}_\text{2D}=L_c(\varepsilon_\perp-1)/4\pi$ for the 2D susceptibility of WTe$_2$ with $L_c=c/2=7.035$ \AA\, ($c=14.07$ \AA\, is the size of the unit cell of the bulk sample along the $c$-axis \cite{Al}) and $\varepsilon_\perp=15.2$ \cite{Kum2}. The binding energy of an exciton with increasing $N_h$ slightly increases (all levels shift down in energy) due to an increase in the effective mass of the hole $m^*_v$ and, as a consequence, an increase in the reduced mass $\mu^*$.

Unlike the usual Coulomb potential $-\widetilde{e}^2/\rho$, a characteristic feature of the exciton energy spectrum is lifting of degeneracy by the angular momentum $l$. The levels shift down in energy from the $s$ level with increasing $l$. With increasing $n$, the splitting level by $l$ decreases.

\begin{table}[b!]
\caption{\label{t1} Calculated values of the binding energies $E_{nl}$, of the variation parameter $a$, and of the average electron-hole distance $\langle\rho\rangle_{nl}$ for the $n=1-3$ states of the exciton with $N_e=N_h=1$ (the substrate is the SiO$_2$ plate with $\varepsilon=3.9$).}
\begin{ruledtabular}
\begin{tabular}{cccc}
State&$E_{nl}$ (meV)&$a$ (\AA)& $\langle\rho\rangle_{nl}$ (\AA)\\
\hline
1$s$&$-$217.88&17.67&17.67\\
2$s$&$-$68.94&24.41&56.96\\
2$p$&$-$85.26&22.48&44.97\\
3$s$&$-$32.80&32.62&123.97\\
3$p$&$-$37.32&30.99&111.57\\
3$d$&$-$42.10&28.56&85.69\\
\end{tabular}
\end{ruledtabular}
\end{table}

\begin{figure}[t!]
\begin{center}
\includegraphics[width=0.48\textwidth]{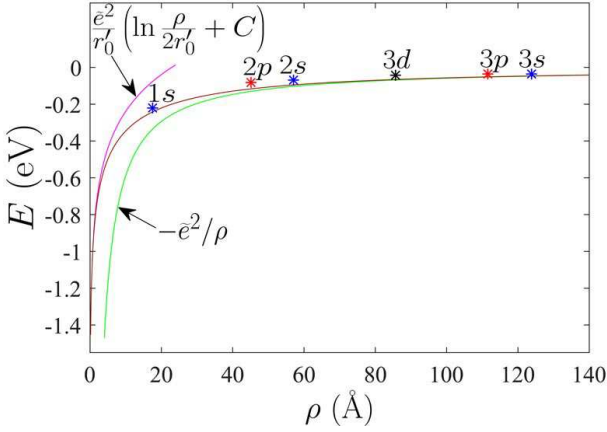}
\caption{\label{f6} (Color online) An illustration of the dependence of $E_{nl}$ on $\langle\rho\rangle_{nl}$ for the first $n=1-3$ levels of the exciton with $N_e=N_h=1$. The blue stars and the red stars correspond to the $ns$ ($n=1,\,2,\,3$) and $np$ ($n=2,\,3$) states, respectively. The black star corresponds to the $3d$ state. Their position in energy is the binding energy in the corresponding state $E_{nl}$, and their position in the coordinate is determined by the average distance $\langle\rho\rangle_{nl}$ (their numerical values are given in Table~\ref{t1}). The brown curve shows potential (\ref{25}), which has two asymptotics: at small distances (the magenta curve) and at large distances (the green curve). The constant $C=0.5772\ldots$ is the Euler constant and $r^\prime_0=20.39$ \AA.}
\end{center}
\end{figure}

It is interesting to note that the exciton levels splitting over $l$ at large $n$ turns out to be small, of the order magnitude of the Coulomb potential change over a large average distance between electron and hole $\langle\rho\rangle_{nl}$. This is illustratively demonstrated in Fig.~\ref{f6} for the exciton with $N_e=1$ and $N_h=1$ (for two other excitons, the picture is qualitatively the same). The energies $E_{nl}$, values of the variation parameter $a$, and the average electron-hole distances $\langle\rho\rangle_{nl}$ are presented in the Table~\ref{t1}. The average distance $\langle\rho\rangle_{nl}$ calculated with using the trial wave functions (\ref{26}) is proportional to the value of the variational parameter $a$, which corresponds to the minimum energy $E_{nl}(a)$: $\langle\rho\rangle_{1s}=a$, $\langle\rho\rangle_{2s}=\hspace{-0.05cm}^7\hspace{-0.08cm}/\hspace{-0.065cm}_3a$, $\langle\rho\rangle_{2p}=2a$, $\langle\rho\rangle_{3s}=\hspace{-0.05cm}^{19}\hspace{-0.08cm}/\hspace{-0.065cm}_5a$, $\langle\rho\rangle_{3p}=\hspace{-0.05cm}^{18}\hspace{-0.08cm}/\hspace{-0.065cm}_5a$, $\langle\rho\rangle_{3d}=3a$ [$a$ is different for each state].

It is worth noting that the 2$s$ state has a slightly larger $\langle\rho\rangle_{2s}$ than the QW width $d$, while the states with $n=3$ noticeably exceed $d$ (in 1.5 times at least). Nevertheless, we find that the quasi-2D consideration of excitons is applicable in this case, although the $n=3$ states lie in the intermediate region between the quasi-2D and quasi-1D behavior of excitons.

Often, starting with $n=3$, the exciton levels ``fall'' on the Rydberg series \cite{Che}, since the potential (\ref{25}) approaches the usual Coulomb potential with a good accuracy (both its asymptotics at small and at large distances are also shown in Fig.~\ref{f6}). In the quasi-2D case, the Rydberg series is \cite{Dur}
\begin{equation}\label{31}
E^\text{(2D)}_n=-\frac{\mu^*\widetilde{e}^4}{2(n-\hspace{-0.05cm}^1\hspace{-0.08cm}/\hspace{-0.065cm}_2)^2}.
\end{equation}
However, for highly excited states, when $\langle\rho\rangle_{nl}\gg d$, we have the quasi-1D behavior of excitons. As is known, the spectrum of the excited exciton states in this case coincides with the spectrum of a three-dimensional (3D) exciton \cite{Bab}
\begin{equation}\label{32}
E^\text{(1D)}_n=-\frac{\mu^*\widetilde{e}^4}{2n^2}.
\end{equation}
Therefore, in the intermediate region between the quasi-2D and quasi-1D behavior of excitons, when $\langle\rho\rangle_{nl}\gtrsim d$, the energies $E_{nl}$ lie between the energies (\ref{31}) and (\ref{32}), $E^\text{(2D)}_n\lesssim E_{nl}\lesssim E^\text{(1D)}_n$.

\subsection{\label{s4b}Strong dielectric screening}

If there is an environment with a large dielectric constant, we obtain that the average electron-hole distance turns out to be much larger than the QW width. Then the behavior of the exciton will be quasi-1D, starting from the ground state. However, the energy of the ground state of an exciton has a logarithmic divergence at short distances in the 1D case \cite{Lou}. To avoid this divergence in our quasi-1D case, we need to take into account that there is a finite scale across the 1D motion, i.e., the presence of the nonzero QW width $d$, and enter the cutoff parameter of the Coulomb potential $d_0\lesssim d$.

On the other hand, the potential (\ref{25}) at large distances transforms into the usual Coulomb potential (see also Fig.~\ref{f5}). Therefore, we can solve the 1D Coulomb problem with a potential that depends only on the relative coordinates of the electron and hole along the QW boundaries (here, along the $y$ axis), where the cutoff parameter $d_0$ is introduced,
\begin{equation}\label{33}
\widehat{U}^\text{(1D)}=\begin{cases}-\widetilde{e}^2/d_0 & \text{for $|y|<d_0$,} \\
-\widetilde{e}^2/|y| & \text{for $|y|>d_0$.}\end{cases}
\end{equation}

The operator of the kinetic energy of the relative 1D motion of the electron and hole is
\begin{equation}\label{34}
\widehat{T}^\text{(1D)}=-\frac{1}{2\mu^*}\frac{\partial^2}{\partial y^2}
\end{equation}
with the same reduced mass $\mu^*$ as above.

As a trial wave function of the ground state, we take
\begin{equation}\label{35}
\widetilde{\psi}_0(y)=\frac{1}{\sqrt{a_0}}\exp\left(-\frac{|y|}{a_0}\right),
\end{equation}
where the variational parameter $a_0$ plays the role of the ground-state Bohr radius.

Averaging Hamiltonian $\widehat{H}^\text{(1D)}_\text{ex}=\widehat{T}^\text{(1D)}+\widehat{U}^\text{(1D)}$ over the ground-state trial wave function (\ref{35}), we express the ground-state exciton energy as \cite{RS}
\begin{equation}\label{36}
E_0=\frac{1}{2\mu^*a^2_0}-\frac{2\widetilde{e}^2}{a_0}\ln\frac{a_0}{d}.
\end{equation}
Here, we do not distinguish between $d$ and $d_0$, since we first carry out the calculation with a logarithmic accuracy.

Minimizing (\ref{36}) with respect to $a_0$, we obtain an equation for $a_0$
\begin{equation}\label{37}
a_0=\frac{a_1}{2\left[\ln\left(a_0/d\right)-1\right]}.
\end{equation}
To the logarithmic accuracy, $\ln(a_1/d)\gg1$, we find the relations
\begin{eqnarray}
E_0&=&-2\mu^*\widetilde{e}^4\ln^2\left(a_1/d\right),\label{38}\\
a_0&=&\frac{a_1}{2\ln\left(a_1/d\right)}.\label{39}
\end{eqnarray}

When $\ln(a_1/d)\sim1$, a more accurate variational calculation should be performed using the modified Coulomb potential \cite{RS}
\begin{equation}\label{33p}
\widehat{U}^\text{(1D)}_\text{m}=-\frac{\widetilde{e}^2}{\sqrt{y^2+d^2_0}}.\tag{34$'$}
\end{equation}
We average the Hamiltonian with potential $\widehat{U}^\text{(1D)}_\text{m}$ over trial function (\ref{35}) to
obtain
\begin{equation}\label{40}
E_0=\frac{1}{2\mu^*a^2_0}-\frac{\pi\widetilde{e}^2}{a_0}\left[H_0\left(\frac{2d_0}{a_0}\right)-Y_0\left(\frac{2d_0}{a_0}\right)\right],
\end{equation}
where $H_0$ and $Y_0$ are the same functions as in the subsection \ref{s4a}, i.e., the average potential energy in Eq.~(\ref{40}) is given by the value of the potential (\ref{25}) at the point $\rho=d_0$ with accuracy to the replacement $r^\prime_0\rightarrow a_0/2$.

Minimizing (\ref{40}) with respect to $a_0$, we obtain an equation for $a_0$
\begin{equation}\label{41}
\begin{split}
&\frac{\pi a_0}{a_1}\left[H_0\left(\frac{2d_0}{a_0}\right)-Y_0\left(\frac{2d_0}{a_0}\right)\right]\\
+&\frac{4d_0}{a_1}\left(1-\frac{\pi}{2}\left[H_1\left(\frac{2d_0}{a_0}\right)-Y_1\left(\frac{2d_0}{a_0}\right)\right]\right)=1.
\end{split}
\end{equation}

The numerical value of the parameter $d_0$ is chosen so that the result obtained by solving equation (\ref{41}) coincides with the result (\ref{38}) for large $\ln(a_1/d)$.

The energy spectrum of excited states ($n=1,\,2,\,3,\ldots$) is given by the formula (\ref{32}), and the Bohr radii are $a_n=na_1$ with $a_1=1/\mu^*\widetilde{e}^2$ \cite{Bab}.

We calculated also the average electron-hole distances for the ground state and the first three excited states: $\langle|y|\rangle_0=\hspace{-0.05cm}^1\hspace{-0.08cm}/\hspace{-0.065cm}_2a_0$, $\langle|y|\rangle_1=\hspace{-0.05cm}^3\hspace{-0.08cm}/\hspace{-0.065cm}_2a_1$, $\langle|y|\rangle_2=3a_2=6a_1$, and $\langle|y|\rangle_3=\hspace{-0.05cm}^9\hspace{-0.08cm}/\hspace{-0.065cm}_2a_3=\hspace{-0.05cm}^{27}\hspace{-0.08cm}/\hspace{-0.065cm}_2a_1$.
For the ground state, we used the wave function (\ref{35}), and for the excited states we took wave functions as eigen wave functions of the Coulomb problem with the potential $-\widetilde{e}^2/|y|$ \cite{RS}
\begin{equation}\label{42}
\psi_n(y)=\frac{\sgn(y)}{\sqrt{2a_n}}\exp\left(-\frac{|y|}{a_n}\right)L^{-1}_n\left(\frac{2|y|}{a_n}\right),
\end{equation}
where $L^{-1}_n$ are the associated Laguerre polynomials.

The numerical values of the energy $E_n$, the Bohr radius~$a_n$, and the average electron-hole distance $\langle|y|\rangle_n$ for the $n=0-3$ states of the exciton with $N_e=N_h=1$ are presented in the Table~\ref{t2}. The system is placed on the TiO$_2$ substrate with $\varepsilon=80$ \cite{Rob}. The ground state energy was calculated with using of Eq.~(\ref{41}), since $\ln(a_1/d)\approx1$ [$a_1=156.1$ \AA\, and $d=52.8$ \AA]. Here, we took $d_0=d$.

\begin{table}[h!]
\caption{\label{t2} Calculated values of the binding energies $E_n$, of the variation parameter $a_n$, and of the average electron-hole distance $\langle|y|\rangle_n$ for the $n=0-3$ states of the exciton with $N_e=N_h=1$ (the substrate is the TiO$_2$ plate with $\varepsilon=80$).}
\begin{ruledtabular}
\begin{tabular}{cccc}
$n$&$E_n$ (meV)&$a_n$ (\AA)& $\langle|y|\rangle_n$ (\AA)\\
\hline
0&$-$3.40&183.33&92.17\\
1&$-$1.14&156.10&234.16\\
2&$-$0.57&312.21&936.63\\
3&$-$0.38&468.31&2107.42\\
\end{tabular}
\end{ruledtabular}
\end{table}

\section{\label{s5}Discussion and Conclusions}

Let us discuss now the possible methods to manufacture the heterostructures under consideration. We assume that it will be necessary to combine the method of applying masks followed by annealing with inert gas ions (argon is often used) and molecular beam epitaxy (MBE). We describe possible technological steps in the Supplementary Material [30]. Annealing is necessary for ``cutting out'' the necessary elements on the TMD monolayer, and MBE is for ``overgrowing'' of the areas subjected to annealing. Recently, monolayers of MoSe$_2$, WSe$_2$, HfSe$_2$, and MoTe$_2$ were grown with the help of MBE \cite{Jiao, Liu1, Yue, Roy}. The mask technique was demonstrated by the example of the synthesis of planar heterostructures based on graphene and hexagonal boron nitride \cite{Liu2}. Thus, we believe that it would be possible to manufacture the proposed and considered here theoretically planar MoTe$_2$/WTe$_2$/MoTe$_2$ QW.

To conclude, the problem of the size quantization of the charge carriers energy levels in such QW is solved both in the two-band and in the four-band approximations, although the latter was actually reduced to a single-band approximation, but taking into account the nearest bands. In particular, the initial effective masses in the conduction band $m^*_{ci}$ and in the valence band $m^*_{vi}$ for the QW regions ($i=1,\,2,\,3$) are considered to be not equal and are taken from the density functional theory calculations. We calculated the effective masses of electrons and holes in the vicinity of the extremes of the dispersion curves corresponding to the size quantization levels.

Using the results for the effective masses, we considered the excitons in the planar QW based on the TMDs monolayers. We proved that there are two regimes of exciton formation, with the weak and strong dielectric screening of the Coulomb potential by the environment.

The former regime is characterized by the quasi-2D behavior of excitons in the ground state and for the first few excited states. Highly excited states in this case fall into the intermediate region between quasi-2D and quasi-1D behavior. The binding energy is calculated using the variational approach. The 2D hydrogen atom eigenfunctions are chosen as the trial wave functions. The latter regime is characterized by the quasi-1D exciton behavior.

The exciton binding energy in 1D case has a logarithmic divergence. To avoid this divergence, we used a modified Coulomb potential, taking into account the finite QW width. The energy of the ground state of the exciton was calculated variaionally. The energy spectrum of the excited states coincides with that of the 3D exciton.

The degeneracy is removed by the angular momentum $l$ in the quasi-2D regime, and the splitting off of the levels occurs down the energy with increasing $l$. This splitting decreases when the principal quantum number increase.

\begin{acknowledgments}
The author is grateful to S.G. Tikhodeev for the helpful discussion and valuable advice on this publication. The work was supported by the Foundation for the Advancement of Theoretical Physics and Mathematics ``BASIS'' (the general formulation of the problem) and by the Russian Science Foundation (the project no. 16-12-10538-$\Pi$, the calculation of the energy spectra of excitons, Sec.~\ref{s4}).
\end{acknowledgments}

\nocite{*}
\bibliography{litExcitQW}
\end{document}


\begin{center}
\large{\textbf{Supplemental Material for:\\ Excitons in planar quantum wells based on transition metal dichalcogenides}}

\vspace{0.25cm}

\normalsize

Pavel V. Ratnikov

\vspace{0.15cm}

\emph{A.M. Prokhorov General Physics Institute, Russian Academy of Sciences, ul. Vavilova 38, 119991 Moscow, Russia}
\end{center}

The growth of monolayer planar heterostructures with a well-controlled physical dimension, quality interfaces and chemical composition is currently a very serious problem. To study excitons in such heterostructures, this problem will have to be overcome. We illustrate in Fig. \ref{SMf1} separate stages of the planar heterostructure MoTe$_2$/WTe$_2$/MoTe$_2$ growth using approach combining molecular beam epitaxy (MBE) and the method of applying and removing masks.

Firstly, it is necessary to select and prepare a substrate for growing the lower one-atom thick layer of chalcogen atoms by MBE (see Fig. \ref{SMf1}a). We propose to use highly ordered pyrolytic graphite (HOPG) as the substrate as in the work \textsl{L.~Jiao et al. New J. Phys. \textbf{17}, 053023 (2015)}. We also believe that the grown heterostructure can be transferred to any other suitable substrate.

Then, we form a mask in the region of the quantum well as a strip of the desired width on the first layer of chalcogen atoms. One-atom thick layer of transition metal atoms is deposited by MBE in the quantum barrier areas (Fig. \ref{SMf1}b).

Thirdly, we apply the second mask in the quantum barriers areas, and remove the first mask in the region of the quantum well. One-atom thick layer of transition metal atoms is deposited by MBE in this region (Fig. \ref{SMf1}c).

Finally, we remove the second mask and apply the top layer of chalcogen atoms (Fig. \ref{SMf1}d).

We have only outlined here in general lines the growth process of the monomolecular thickness planar heterostructures based on transition metal dichalcogenides. We are aware that there are many technological obstacles in this way that are likely to have a significant impact on the feasibility of synthesizing these heterostructures. In particular, the process of the mask forming can be fundamentally carried out inside the MBE reactor, while the removing the masks will require exposure to chemicals (solvents) at higher pressures. Therefore, the heterostructure will have to be removed from the ultra-high vacuum region inside the MBE reactor to the high-pressure compartment, followed by return to the reactor.

\begin{figure}[b!]
\begin{center}
\includegraphics[width=0.45\textwidth]{fig1a}\hspace{0.075\textwidth}\includegraphics[width=0.45\textwidth]{fig1b}

\vspace{0.05\textheight}

\includegraphics[width=0.45\textwidth]{fig1c}\hspace{0.075\textwidth}\includegraphics[width=0.45\textwidth]{fig1d}
\caption{\label{SMf1} (Color online) Consecutive stages of the growth process for the planar heterostructure MoTe$_2$/WTe$_2$/MoTe$_2$: \textbf{(a)}~a~deposition of a one-atom thick layer of tellurium atoms on HOPG, \textbf{(b)} the subsequent deposition of molybdenum atoms on this one-atom thick layer with a pre-applied mask (Mask 1), \textbf{(c)} applying another mask (Mask 2) on the sides of Mask 1, removing Mask 1 and subsequent deposition of tungsten atoms, \textbf{(d)} removing Mask 2 and subsequent deposition of the second one-atom thick layer of tellurium atoms.}
\end{center}
\end{figure}